\documentclass[11pt]{article}
\usepackage{mymoriond,epsfig}

\begin{document}

\begin{flushright}
CLNS~02/1796\\
{\tt hep-ph/0207357}\\[0.2cm]
July 30, 2002
\end{flushright}

\vspace*{2cm}
\title{MORIOND QCD 2002: THEORETICAL SUMMARY}

\author{M. NEUBERT\footnote{Supported by the National Science Foundation 
under Grant PHY-0098631}}

\address{F.R. Newman Laboratory for Elementary-Particle Physics\\
Cornell University, Ithaca, NY 14853, USA}

\maketitle\abstracts{I summarize highlights of the theory talks presented
at the 37th Rencontres de Moriond on QCD and High Energy Hadronic 
Interactions.}

\section{Preface}

During this conference we heard many impressive talks on a vast variety 
of subjects. Although QCD is a mature field, very significant progress is 
still being made. Understanding QCD is not just an academic challenge but
impacts on almost all aspects of high-energy physics. In particular, it is
a prerequisite for precise measurements of many Standard Model parameters,
such as the gauge couplings $\alpha_s$ and $\alpha$, fermion and boson 
masses, flavor-changing couplings, and the CP-violating phase of the CKM 
matrix. Understanding QCD is also important for searches for physics 
beyond the Standard Model, both via the direct production of new 
particles and using precision measurements at low energy. Last but not 
least, QCD is our playground for exploring strongly coupled gauge 
theories; lessons learned here will help in understanding other strongly 
coupled theories (i.e., almost any New Physics model, perhaps even 
gravity).

In this talk I will focus on three sectors of QCD research: core QCD 
(pQCD, resummation, power corrections, factorization), multi-body 
problems in QCD (saturation, unitarization, heavy ions), and searches for
New Physics and CP violation. Rather than repeating all 39 theory talks 
of this conference, I will try to put things in perspective and focus on 
a few recent developments. I apologize to all those whose interesting 
contributions are only briefly mentioned here because of space (time) 
limitations.

\section{Hard-Core QCD (rated {\em R})}

\begin{figure}[ht]
\epsfxsize=14cm
\centerline{\epsffile{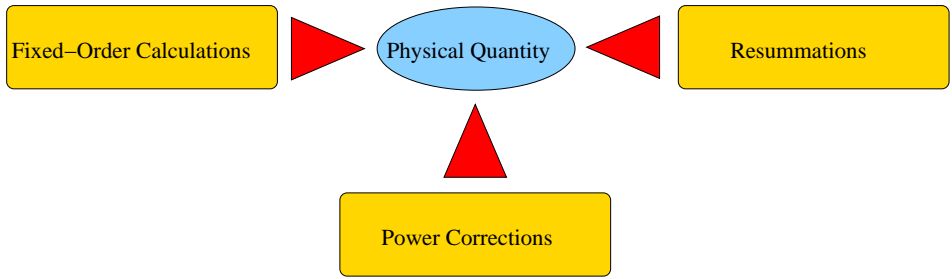}}
\vspace{-0.5cm}
\caption{\label{fig:trinity}
Multi-loop amplitudes, resummation of large logarithms, and 
non-perturbative power corrections form the basis of modern QCD 
calculations.}
\end{figure}

Precise calculations of physical cross sections in QCD pose several 
theoretical challenges. Recently, significant progress has been made
on various fronts, pushing the limits of what is state-of-the-art to
a new level of sophistication. Research in this area has focused on 
three directions: multi-loop amplitudes, resummation of large 
logarithms, and non-perturbative power corrections. Although rather
different in their technicalities, these three major directions go
together and form the basis of most precision calculations in QCD, as
indicated in Figure~\ref{fig:trinity}.

\subsection{Scattering Amplitudes Beyond Leading Order}

Controlling the scale dependence (and, more generally, the dependence on 
the choice of the renormalization scheme) in the prediction for a QCD 
amplitude requires that the calculations be performed beyond the leading 
order. This requires exact multi-loop calculations of Feynman diagrams. 
There has been impressive progress in recent years in the calculation of 
multi-parton scattering amplitudes at next-to-leading order (NLO) and 
beyond. Examples include jets in $e^+ e^-$ and hadron collisions, Higgs 
production, Drell--Yan production, etc. The challenges faced in such 
calculations were discussed by Oleari. They are the evaluation of (many) 
two-loop diagrams, the need for next-to-next-to-leading order (NNLO) 
splitting functions, the handling of real emissions (soft and collinear 
cancellations), and finally the implementation of hadronization effects 
using Monte Carlo generators, also discussed by Weinzierl.

An important example of a NNLO calculation is the inclusive Higgs 
production at hadron colliders, which was reviewed by Kilgore and 
Grazzini. The cross section for this process increases strongly when 
going from leading to NLO, and so a NNLO calculation is required to see 
whether or not a reliable prediction can be obtained. Such a NNLO 
calculation becomes feasible after introducing an effective vertex for 
the $Hgg$ coupling (with the top loop integrated out). The result shows 
that perturbation theory converges better than expected based on 
previously available partial resummations. Fortunately, it appears that
there is now a reliable prediction for this important discovery process.

\subsection{Resummations}

While exact multi-loop calculations are indispensable for obtaining 
precise predictions, in many cases they are insufficient due to the 
presence of widely separated mass scales. Such scales typically arise 
when experimental cuts restrict phase space, or when heavy particles are 
involved. Physical quantities are infrared safe, but large logarithms can
arise as a result of incomplete infrared cancellations near phase-space 
boundaries. Such logarithms lead to a breakdown of fixed-order 
perturbation theory, making it necessary to resum an infinite number of 
terms in the perturbation series. While the leading double logarithms are
under good control, Salam pointed out that the resummation of the NLO 
single logarithms still poses significant theoretical challenges. As an 
example he discussed so-called ``non-global'' observables, which measure 
gluon emission only in part of the event. In this case the approximation 
of independent emissions implies suppression only of primary emissions in 
the current hemisphere. But for a correct resummation of single logarithms 
one must also suppress energy-ordered large-angle secondary emissions. 
Accounting for the non-global logarithms thus needs a change of philosophy.

A semi-numerical method for the computation of single-logarithmic 
effects due to multiple gluon radiation was suggested by Zanderighi, who 
presented new predictions for several event-shape distributions (thrust 
major, oblateness, 3-jet resolution). Novel effects can also arise from 
the interplay of different types of logarithms. Kulesza explained how the
simultaneous resummation of recoil and threshold logarithms in
electroweak boson production leads to an interplay of Sudakov suppression 
and enhancement.

\subsection{Power Corrections}

\begin{figure}[ht]
\epsfxsize=14cm
\centerline{\epsffile{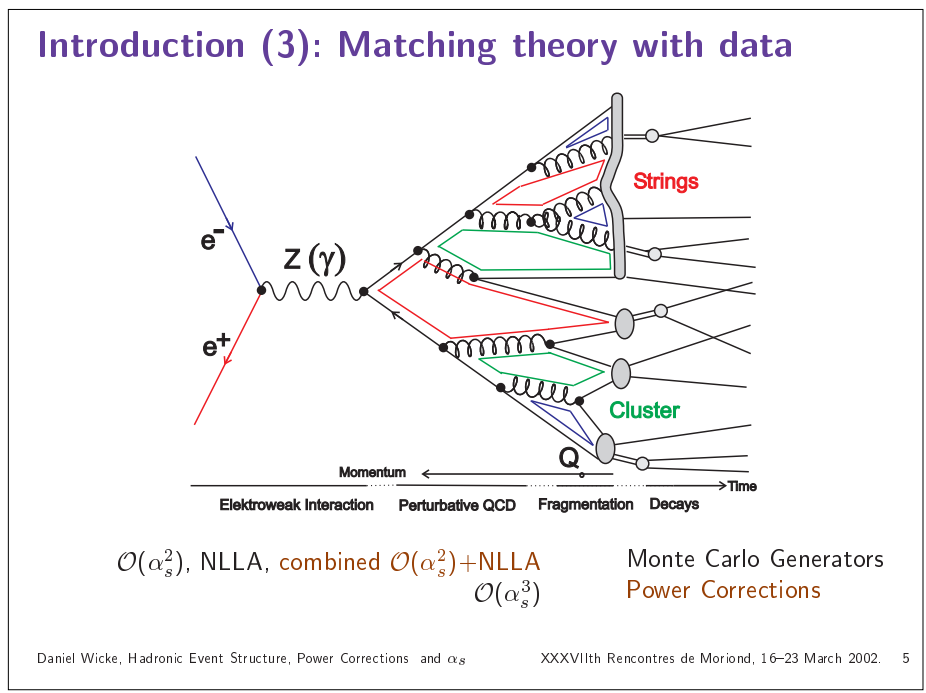}}
\vspace{-0.2cm}
\caption{\label{fig:hadrons}
An artistic view of hadronization effects, taken from Wicke's talk.}
\end{figure}

Unfortunately, physics doesn't stop with partons, and the presence of 
hadronization effects complicates our understanding of QCD observables, 
as indicated in Figure~\ref{fig:hadrons}. This is what adds the spice to 
the life of QCD phenomenologists.

There are contributions to scattering amplitudes not seen at any (finite)
order in perturbation theory. Such non-perturbative effects are often 
power suppressed by a large scale $\sim(\Lambda/Q)^n$. In simple cases, 
they can be included using the operator product expansion (OPE), which 
provides a systematic parameterization of power corrections in terms of 
matrix elements of local operators. Examples include moments of structure
functions discussed by Alekhin, and a gauge-dependent 
$\langle A^2\rangle/Q^2$ correction in lattice determinations of 
$\alpha_s$ mentioned by Quintero, which is a lattice artifact that must 
be subtracted before taking the continuum limit.

There is evidence for an interplay of perturbative terms and power 
corrections in the sense that often power corrections are found to 
diminish if higher-order perturbative corrections are included. In 
general, however, power corrections cannot be made arbitrarily small, and
an estimate of their effects is often the limiting factor in theoretical 
predictions. Much effort is currently being devoted to more systematic 
studies of power corrections in cases where no local OPE can be applied.
For many years, the method of choice was based on the renormalon 
calculus, by means of which one can study the renormalization-group 
mixing of (certain) terms of different power in $1/Q$. This method has 
been applied rather successfully to event shapes in $e^+ e^-$ 
annihilation, which as Marchesini reviewed can be fitted (within large
errors) in terms of only two parameters: $\alpha_s(M_Z)$ and 
$\alpha_0(\mu_I)$. Both Marchesini and Salam stressed, however, that 
event shapes in deeply inelastic scattering (DIS) are not just a simple 
extension of $e^+ e^-$ but pose additional problems. For instance, the 
question about the universality (i.e., process independence) of the 
hadronic parameter $\alpha_0(\mu_I)$ arises. Another interesting 
application, presented by Qiu, employs the $b$-space method of Collins, 
Soper and Sterman to estimate the non-perturbative contribution to 
resummation formulae for heavy boson production, using a new 
extrapolation to the large-$b$ region.

More recently, significant progress has been made toward a systematic 
analysis of power corrections for observables that do not admit an 
expansion in local operators. Factorization theorems provide a separation 
of different energy scales (hard, collinear, soft, ultrasoft) to all 
orders of perturbation theory. Traditionally, they are derived by an 
analysis of Feynman diagrams using the method of regions. The example of 
factorization in heavy-quark fragmentation was discussed by Cacciari and 
Corcella. Another example (which I would have discussed if I had not been 
convinced to give the summary talk) is the QCD factorization approach to 
hadronic $B$-meson decays developed by Beneke, Buchalla, Sachrajda and 
myself. Alternatively, factorization theorems can be derived by 
constructing effective field theories for soft and collinear fields. 
Examples in the context of non-relativistic QCD were presented by Zhang 
and Vairo. Although not discussed at this conference, I consider the 
recent development of the soft-collinear effective theory by Bauer, Pirjol 
and Stewart a significant step forward. This theory has potential 
applications in many areas of QCD phenomenology. Another interesting 
development was presented by Gardi, who made the conjecture of a 
factorization formula for the moments of the structure function 
$F_2(x,Q^2)$, which is believed to be valid beyond leading power.

\subsection{Other Topics}

Let me finish this section by mentioning several other interesting talks,
which were not related to perturbative QCD. Igi presented a new look at 
an old subject, the large-$s$ behavior of the $\pi$--$N$ cross section, 
using finite-energy sum rules. Semi-classical quantization of effective 
string theories and Regge trajectories were studied in great detail by 
Baker. Tung presented a new generation of parton distribution functions, 
with uncertainties from the global QCD fit taken into account. New 
results on Fierz transformations and bosonization were discussed by 
J\"ackel, while Arleo summarized constraints on quark energy loss from 
Drell--Yan data.

\section{QCD in Many-Body Systems (rated {\em PG-13})}

While exact perturbative calculations are only possible for very simple
processes involving few partons, new challenges are met when one attempts 
to understand the properties of hadronic, nuclear or quark matter.

\subsection{Non-perturbative Effects on Structure Functions}

Kulagin emphasized that the structure functions of nuclei are not simply 
multiples of nucleon structure functions, but that several novel effects
must be taken into account. These are, in particular, the shadowing 
effect at small $x$, nuclear binding and off-shell corrections at 
intermediate $x$, and nuclear motion (Fermi motion) at large $x$. 
Constraints on nuclear gluon shadowing obtained from DIS data were 
pointed out by Salgado. 

Liuti argued that, at large $x$ and low $Q^2$ ($W^2<2.5$\,GeV$^2$), 
structure functions can no longer be described by standard pQCD evolution,
but instead exhibit significant scaling and duality violations. A 
``semi-hard'' cluster mass distribution was introduced, describing the 
rescattering of the proton remnant 
($p\to\mbox{cluster}\to\mbox{partons}$).

Very interesting effects occur at low $x$ and/or very large $A$, where 
saturation and unitarization corrections become important. As Iancu 
pointed out, parton distributions rise strongly at small $x$. A linear 
evolution \`a la BFKL or DGLAP can explain the rise but leads to 
inconsistencies such as violation of unitarity. At high density, 
non-linear effects should limit the growth.

Capella, Kaidalov, Ferreiro and Salgado have suggested a hybrid approach 
to DIS and diffraction, in which Regge theory is combined with pQCD. 
Properties of this model were discussed by Kaidalov and Ferreiro. 
Unitarity is restored by including multi-pomeron exchange. For instance, 
in $ep$ collisions the virtual photon dissociates into a $(q\bar q)$ 
fluctuation, which for small transverse size is described by pQCD (color 
dipole), but for large size is described by Regge phenomenology. This 
model provides a good description of the HERA data for $F_2(x,Q^2)$. 

An alternative approach discussed by Steffen links gluon saturation to
unitarity in a model based on the functional integral approach, which 
relates the transition amplitude to correlation functions of Wilson loops.
In that way a unified description of $pp$, $\gamma^*p$ and $\gamma\gamma$
reactions can be obtained.

\subsection{The Color-Glass Condensate}

A very interesting formal development discussed by Iancu, Itakura and 
Kharzeev is that of an effective theory, derived from QCD, for (very) 
high-density gluonic systems at (very) small $x$. Saturation occurs when 
the interaction probability becomes $O(1)$, i.e., for $Q^2<Q_S^2(x)$ with
a saturation scale given by 
$Q_S^2(x)=\alpha_s N_c\cdot(x\,G(x,Q^2)/\pi R_A^2)$. A crucial 
observation, stressed by Iancu, is that 
$Q_S^2(x)\sim A^{1/3} x^{-C\alpha_s}$ with a coefficient $C>0$. It 
follows that the saturation scale becomes perturbative in the formal 
limit where $A\to\infty$ and/or $x\to 0$. One then enters a
semi-classical regime of weak coupling and large occupation numbers.
The high-density gluons correspond to classical color fields in the 
effective theory, which are radiated by fast-moving partons. In other 
words, the fast partons are ``frozen'' in some random color 
configuration, which the authors have called a ``color glass'' (in 
analogy with spin glasses). 

Some predictions of this theory have been discussed by Itakura. The gluon
distribution in transverse phase space saturates for small $k_\perp$ and 
falls off over a region $Q_S<|k_\perp|<Q_S^2/\Lambda$, yielding geometric
scaling, i.e., $\sigma_{\rm tot}^{\gamma^* p}(x,Q^2)\to f(Q^2 R_0^2(x))$. 
Scaling holds over a wide region $0.045<Q^2<450$\,GeV$^2$, since 
$Q_S/\Lambda\sim 5$--20 is a large scale. Predictions for hadron 
production at RHIC (multiplicity and rapidity distributions) were 
presented by Kharzeev.

In the discussion sessions there was much controversy about the 
phenomenological applications of these ideas, mainly related to the fact 
that the shadowing corrections predicted by the color-glass theory are
larger than those seen in the data. The question was raised whether this
might be ``a beautiful theory, which however fails when applied to 
present data''? Future will tell. In my opinion one cannot overemphasize
the importance of having first-principles predictions in some limit of 
QCD, even if this limit is not so close to reality. Therefore, work on 
the color glass theory is certainly worth pursuing.

\subsection{Chiral Phase Transition in $\chi$PT}

The temperature dependence of the chiral condensate, the order parameter 
of chiral symmetry breaking, can be studied in chiral perturbation theory 
using the virial expansion and unitarization, as pointed out by Pelaez. 
He observes a ``paramagnetic effect'' (reduction of the critical 
temperature $T_c$ when going from $n_f=2$ to $n_f=3$ light flavors) and a
``ferromagnetic effect'' (reduction of $T_c$ for $m_q\ne 0$). These 
analytical results are consistent with lattice computations.

\subsection{Other Topics}

There were several other interesting presentations related to heavy-ion 
physics. Pierog discussed the role of screening corrections in numerical 
simulations. Baryon-number transfer was discussed by Shabelski, while 
Sousa presented predictions for the baryon and anti-baryon yields 
obtained in the dual parton model. Finally, Sarcevic presented detailed 
calculations of prompt photon production for RHIC and LHC.

\section{QCD in Flavor Physics and New Physics Searches 
(rated {\em General Audience})}

The last few years have seen a revolution in $B$ physics. At this year's
spring conferences, BaBar and Belle (and CLEO) have presented yet another
round of exciting results, such as updated precision measurements of 
$\sin 2\beta$, measurements of mixing-induced and direct CP violation in 
$B\to\pi^+\pi^-$ decays, measurements of (and limits on) direct CP 
asymmetries in several decay modes, updated results for rare charmless 
and radiative decays, and last but not least new precision determinations 
of the CKM matrix elements $|V_{cb}|$ and $|V_{ub}|$. This wealth of 
experimental information has triggered a steady improvement of the 
theoretical tools that allow us to interpret these data in terms of 
Standard Model parameters. This is non-trivial, because the physics of
hadronic weak decays is to a large extent the physics of hadronic bound 
states.

\subsection{Unitarity Triangle}

\begin{figure}[t]
\centerline{\epsfxsize=8cm\epsffile{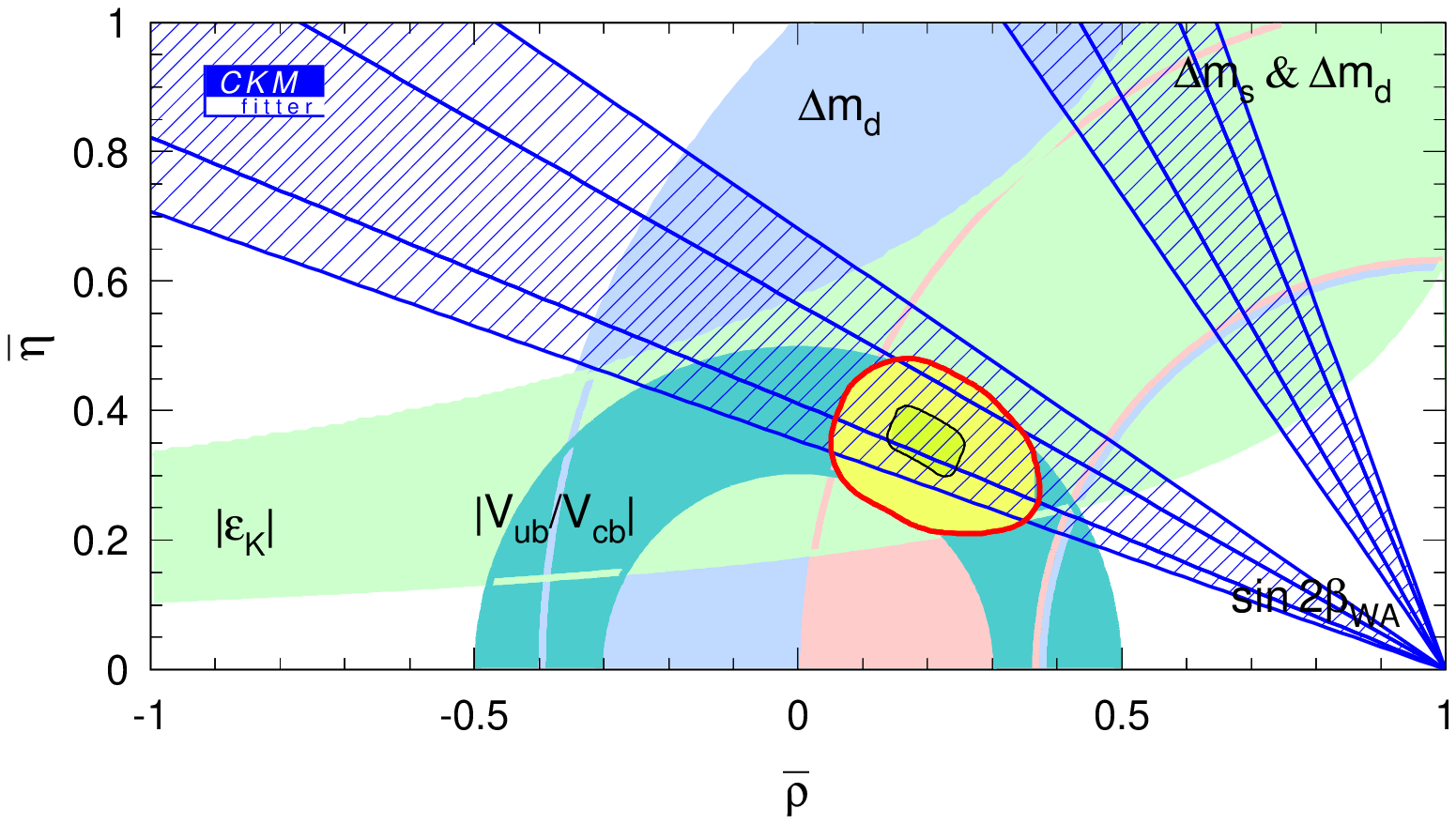}\quad%
\epsfxsize=7cm\epsffile{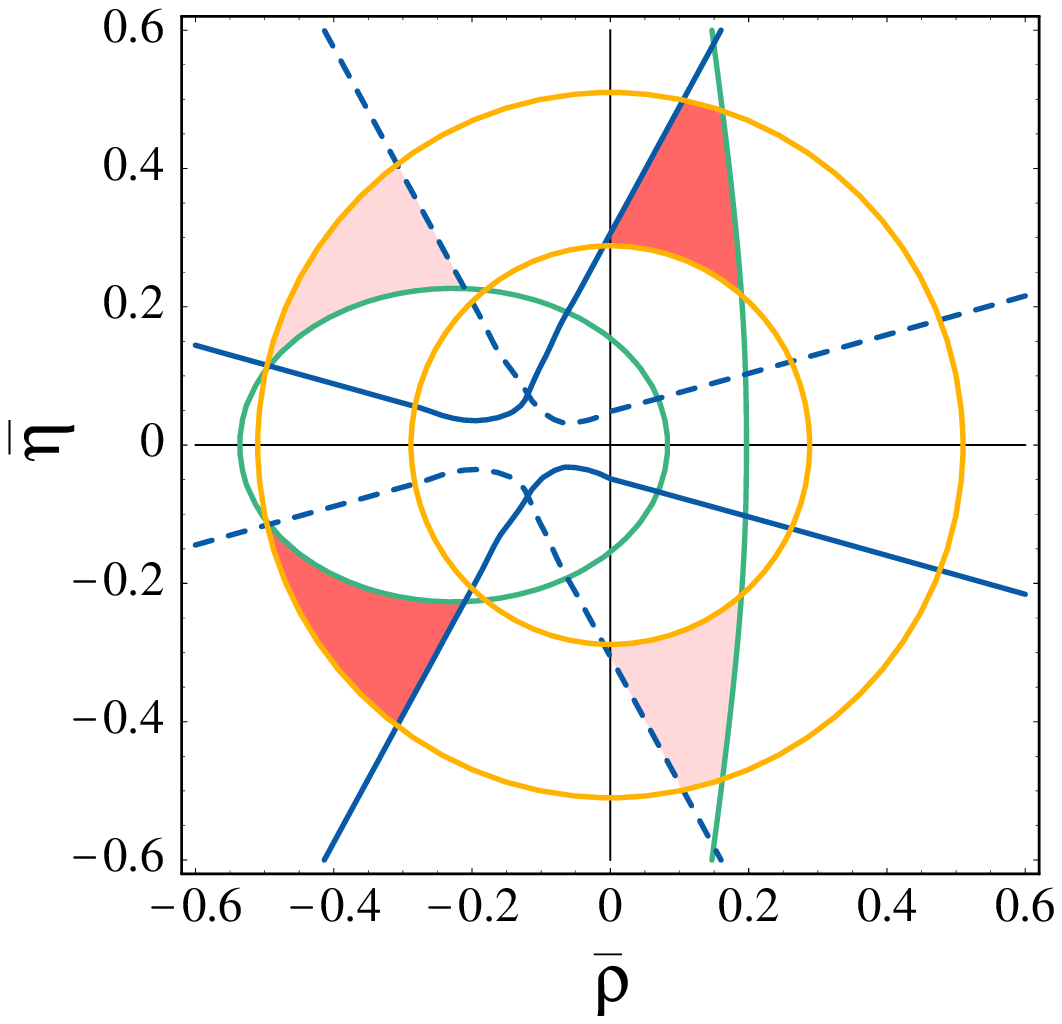}}
\vspace{-0.1cm}
\caption{\label{fig:UTfit}
Left: Standard constraints on the apex $(\bar\rho,\bar\eta)$ of the 
unitarity triangle. Right: Allowed regions in the $(\bar\rho,\bar\eta)$ 
plane obtained from a novel construction of the unitarity triangle.}
\end{figure}

The experimental knowledge about the smallest entries in the CKM matrix
($V_{ub}$ and $V_{td}$) can be summarized by displaying the unitarity 
relation $V_{ub}^*\,V_{ud}+V_{cb}^*\,V_{cd}+V_{tb}^*\,V_{td}=0$ as a 
triangle in the complex $(\bar\rho,\bar\eta)$ plane. As is well known, 
CP violation results from $\bar\eta\ne 0$ and so corresponds to a 
non-vanishing area of the triangle. The so-called ``standard 
constraints'' on the apex of the unitarity triangle come from 
measurements of CP violation in $K$--$\bar K$ mixing (parameter 
$\epsilon_K$), $|V_{ub}/V_{cb}|$ in semileptonic decays of $B$ mesons, 
the neutral $B$-meson mass differences $\Delta m_{d,s}$ in 
$B_{d,s}$--$\bar B_{d,s}$ mixing, and $\sin 2\beta$ in $B\to J/\psi\,K_S$
decays. A summary of the resulting constraints is shown in the first plot
in Figure~\ref{fig:UTfit}. With the exception of the $\sin2\beta$ 
measurement, the standard analysis is limited by large theoretical 
uncertainties, which dominate the widths of the various bands in the 
figure. These uncertainties enter via the calculation of hadronic matrix 
elements. 

With the new data of BaBar and Belle, it is possible to construct the
unitarity triangle using novel methods based on charmless hadronic $B$ 
decays. They are afflicted by smaller hadronic uncertainties and hence 
provide powerful new tests of the Standard Model, which can complement 
the standard analysis. The resulting constraints on $(\bar\rho,\bar\eta)$
are independent of $B$--$\bar B$ and $K$--$\bar K$ mixing. They are in 
this sense orthogonal to the standard analysis. Specifically, one 
combines information from semileptonic $B$ decays, CP-averaged branching 
fractions in $B^\pm\to(\pi K)^\pm$ decays, and the time-dependent CP 
asymmetry in the decays $B\to\pi^+\pi^-$. The result of such an analysis,
using present data, is shown in the second plot in Figure~\ref{fig:UTfit}.

There are four allowed regions, two of which remain if we use the 
information that the measured value of $\epsilon_K$ requires a positive 
value of $\bar\eta$. One of these regions (dark shading) is close to the 
standard fit. This agreement is highly non-trivial, since with the 
exception of $|V_{ub}|$ none of the standard constraints are used in this 
construction. Interestingly, there is a second allowed region (light 
shading) which would be consistent with the constraint from $\epsilon_K$ 
but inconsistent with the constraints derived from $\sin2\beta$ and 
$\Delta m_s/\Delta m_d$. Such a solution would require a significant New 
Physics contribution to $B$--$\bar B$ mixing.

\subsection{Anomalous Magnetic Moment of the Muon}

If you type {\sc ``fits supersymmetry like a glove''} in Google, you find
an article published in the science section of the New York Times on
Feb.~9, 2001, which is entitled {\em Tiniest of Particles Pokes Big Hole 
in Physics Theory}. There we find the following modest statements by some
leading physicists:

\medskip\noindent
{\em ``The most natural meaning of this kind of indication,'' 
Dr.~Marciano said, ``would be supersymmetry.'' The observed change in 
frequency, he said, ``fits supersymmetry like a glove.''}

\medskip\noindent
{\em ``It would mean that in describing the world, we would need to add 
to the equations of the Standard Model,'' Dr.~Wilczek said. ``And those 
additions make the whole thing much prettier, more unified and more 
beautiful.''}

\medskip\noindent
And best of all: {\em ``It could lead to a whole deeper understanding of 
how reality is put together'', Dr.~Gabrielse said.}

\medskip\noindent
Unfortunately, by now the supersymmetrists have once again to find a 
reason why SUSY does {\em not\/} give a sizable contribution to an 
observable, since the $(g_\mu-2)$ anomaly has essentially disappeared 
with a sign mistake! As everybody knows, the contraction of two 
$\epsilon$-tensors is proportional to the determinant of the space-time 
metric: $\epsilon_{\mu\nu\alpha\beta}\,\epsilon^{\mu\nu\alpha\beta}
=24\,\mbox{det}(g^{\mu\nu})$. Since we live in 1 time and 3 spatial 
dimensions (or so we think), this number is $-24$. In {\sc FORM}, this 
quantity is $+24$ (as explained on p.~14 of the tutorial), and there the 
trouble begins \dots (and my polemics stops).

\begin{figure}[ht]
\epsfxsize=10cm
\centerline{\epsffile{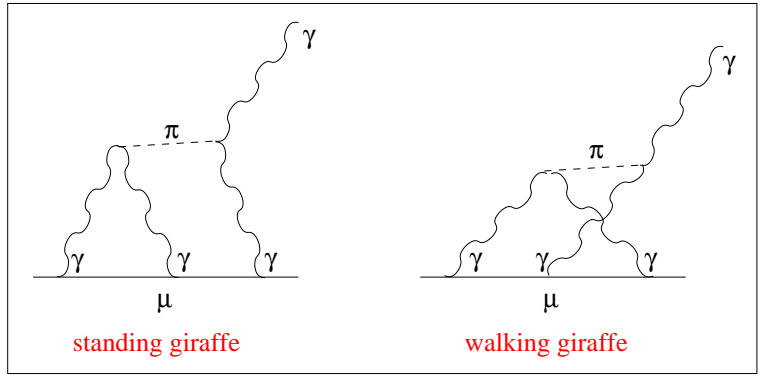}}
\vspace{0.1cm}
\caption{\label{fig:giraffe}
``Giraffe diagrams'' entering the light-by-light scattering contribution
to the anomalous magnetic moment of the muon.}
\end{figure}

What is left after the dust has settled is a hard QCD problem, since the 
main uncertainty in the calculation of $a_\mu=(g_\mu-2)/2$ comes from 
hadronic contributions such as light-by-light scattering, discussed by 
Czarnecki. The relevant diagrams shown in Figure~\ref{fig:giraffe} are 
dominated by soft physics and thus cannot be computed reliably. The 
corresponding uncertainty makes up for a large portion of the difference 
$a_\mu^{\rm exp}-a_\mu^{\rm th}=(202\pm 151_{\rm exp}\pm 100_{\rm th})
\cdot 10^{-11}$. While the double and single chiral logarithms in this 
estimate are sort of under control, the non-logarithmic contribution to 
light-by-light scattering is largely model dependent. To improve the 
situation would require a better control over the $\pi\gamma^*\gamma^*$ 
form factor, which was discussed by Dorokhof and Praszalowicz.

\subsection{Black Holes at Future Colliders}

Even a QCD conference can nowadays not avoid having a talk on extra 
dimension. While his presentation was totally unrelated to QCD, 
Landsberg presented perhaps the most fancy transparencies seen at this 
conference. If his claim that the LHC will turn out to be a black-hole 
factory is correct, this machine should be renamed the {\em Large Hole 
Producer (LHP)}.

\section{Fromages et Desserts}

As at any good meeting, a couple of surprises were presented and lively 
discussed at this conference. I therefore finish with my list of three 
$3\sigma$ effects.

The measurement of the weak mixing angle in deep-inelastic neutrino 
scattering presented by NuTeV, $\sin^2\theta_W=0.2277\pm 0.0016$, 
deviates by $3\sigma$ from the value obtained from the global electroweak
fit. If this discrepancy is real (i.e., if it cannot be explained by some
underestimated hadronic uncertainty), it could be explained in terms of a 
left-handed coupling $g_L$ lower than predicted by the Standard Model, 
whereas the right-handed coupling $g_R$ appears to be about right.

Belle has reported two anomalies at this conference. The first is the
observation of a near maximal direct CP asymmetry in $B\to\pi^+\pi^-$ 
decays, $A_{\rm CP}=0.94_{\,-0.31}^{\,+0.25}\,\pm 0.09$. This is a factor
3 larger than even the most optimistic theoretical predictions. The 
second Belle anomaly is equally surprising. They see a large direct CP 
asymmetry in $B^\pm\to\pi^\pm K_S$ decays, 
$A_{\rm CP}=0.46\pm 0.15\pm 0.02$, which deviates from zero by about
$3\sigma$. However, in the Standard Model these decays are almost pure
penguin processes and so lack the required amplitude interference, which 
could result is a large direct CP asymmetry. The Standard Model 
expectation is $A_{\rm CP}<3\%$.

If only one of these three $3\sigma$ effects will turn out to be real, 
then perhaps Moriond 2002 will be remembered as the conference where the 
Standard Model begun to collapse.

\section*{Acknowledgments}
I wish to thank the organizers of Moriond QCD for the invitation to 
present this summary talk, and many of my colleagues (you know who you 
are) for great times on (and off!) the slopes of Les Arcs.

\end{document}